\documentclass{edp-jp4}
\usepackage{graphicx}

\begin{document}

\title{Non--equilibrium relaxation of an elastic string in random
media} 
\author{Alejandro B. Kolton}\address{Universit\'e de Gen\`eve, DPMC, 24 Quai Ernest Ansermet,
CH-1211 Gen\`eve 4, Switzerland.}
\author{Alberto Rosso}
\address{ Laboratoire de physique th\'{e}orique et
mod\`{e}les statistiques CNRS UMR8626, B\^{a}t. 100 Universit\'{e}
Paris-Sud; 91405 Orsay Cedex, France.}
\author{Thierry Giamarchi}\sameaddress{1}
\maketitle
\begin{abstract}
We study the relaxation of an elastic string in a two
dimensional pinning landscape using Langevin dynamics simulations.
The relaxation of a line, initially flat, is characterized by a
growing length, $L(t)$, separating the equilibrated short length
scales from the flat long distance geometry that keep memory of the
initial condition. We find that, in the long time limit, $L(t)$ has
a non--algebraic growth, consistent with thermally activated jumps
over barriers with power law scaling, $U(L) \sim L^\theta$. 
\end{abstract}

Understanding the dynamics of elastic manifolds in random media has been the
focus of intense activities both on the theoretical and experimental side
\cite{giamarchi_sitges_review}.  Because of the competition between disorder
and elasticity, glassy properties arise leading to divergent barriers
separating metastable states.  The creep, {\it i.e.} the steady--state
response to a small external force, has been studied theoretically
\cite{theory_creep,kolton_creep} and observed experimentally
\cite{experiment_creep}. Much less is known about the non-stationary
relaxation to equilibrium. Theoretical attempts to tackle this problem have
been made using mean field and renormalization group approaches
\cite{cugliandolo_relaxmanifold_balents_tbl,schehr_dynamics}, while
numerical simulations have been mostly restricted to 2--dimensional systems
\cite{numerics_relax}. Direct applications of these results to one
dimensional domain walls are however difficult.

 In this paper we study, numerically, the slow non-stationary motion of an
elastic string in a two dimensional random medium, relaxing from a flat
initial configuration. The string obeys the equation of motion $\gamma
\partial_t u_(z,t)= c \partial_z^2 u(z,t) + F_p(u,z) + \eta(z,t)$, where
$\gamma$ is the friction coefficient and $c$ the elastic constant,
$F_p(u,z)$ the pinning force for a random bond disorder and $\eta(z,t)$ is
the thermal noise. The details of the numerical method are given in
\cite{kolton_relax}.

\begin{figure}
 \centerline{\includegraphics[width=11.0cm,height=6.0cm]{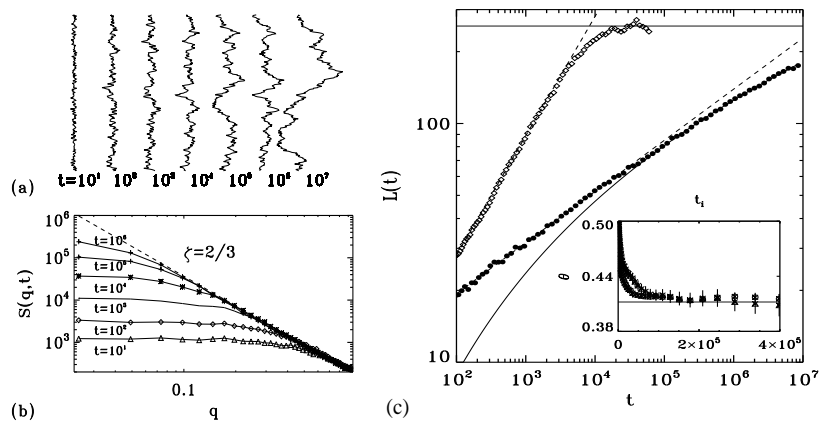}}
 \caption{(a) Typical configurations of the relaxing string (b) $S(q,t)$ at
 $T=0.5$. The dashed line corresponds thermal equilibrium solution. (c)
 Characteristic length scale $L(t)$ of a string of size
 $L=256$. ($\diamond$) symbols and the dotted line correspond to the
 numerical and analytical $L(t)$ for the clean system, and ($\bullet$)
 symbols to the disordered case. The continuum line fits to the logarithmic
 growth, and the dashed line is a fit to a power law growth at intermediate
 scales.  Inset: exponent $\theta$ of the logarithmic law extracted from the
 fit in the time interval $t_i<t<10^6$, for $L=512$ ($\triangle$) and
 $L=256$ ($\circ$).}
 \label{fig:fig1y2}
\end{figure}

Fig.~\ref{fig:fig1y2}(a) shows the typical relaxation of a string and
Fig.~\ref{fig:fig1y2}(b) the evolution of the structure
factor $S(q,t)$ which characterize the geometry of the line
\cite{kolton_creep,kolton_relax}. At short length scales the line has
reached equilibrium and it is characterized by the well known roughness
exponent $\zeta=2/3$ \cite{kardar_exponent_line}. At large length scales a
plateau still keeps memory of the initial flat condition. A unique crossover
length $L(t)$ can be defined. Its evolution is shown in
Fig.~\ref{fig:fig1y2}(c).
Two main scenarios have been proposed to describe the long time 
growth of $L(t)$.  The
first one relies on phenomenological scaling arguments, based on creep. At low
temperatures the relaxation is dominated by the energy barriers $U(L)$ that
must be overcomed in order to equilibrate the system up to a length scale
$L$. Using the Arrhenius thermal activation law we can thus express the
relaxation time $t(L) \sim \exp[\beta U(L)]$.  Even if the exact numerical
determination of $U(L)$ is an NP-complete problem it is usually conjectured
that the typical barriers of the energy landscape scale, asymptotically with
$L$, the same way as the free energy fluctuations: $U(L)\sim L^{\theta}$,
with $\theta=1/3$ for a line. Numerical calculations \cite{drossel_barrier}
and FRG calculations \cite{theory_creep} seem to confirm this conjecture.
Following these arguments we infer that 
$L(t) \sim [\log(t)]^{\frac{1}{\theta}}$
\cite{kolton_relax,yoshino_unpublished}.  The second scenario predicts
$L(t)\sim t^{1/z}$, like for the clean system but with a new temperature
dependent exponent $z>2$ \cite{schehr_dynamics,numerics_relax}.  This power
law scaling can be ruled out from our data, due the observed bending in the
log-log scale in Fig.~\ref{fig:fig1y2}(c). On the other hand a
two-parameters fit to the logarithmic growth law gives an exponent $\theta$
which, at long times, becomes size and time independent, as shown in the
inset of Fig.~\ref{fig:fig1y2}. However we found $\theta \approx 0.49$. This
value, bigger than the expected $\theta=1/3$, indicates either a violation of
the expected scaling of barriers or the presence of logarithmic corrections
in the leading term:  $U(L)\sim L^{1/3} \log^{\mu}(L)$
\cite{drossel_barrier,yoshino_unpublished}.


\end{document}